\documentclass{article}
\begin{document}
\title{Computation of the spectra of the quasars.}
\author{Jacques Moret-Bailly 
\footnote{Laboratoire de physique, Université de Bourgogne, BP 47870, F-21078 Dijon cedex, France.
Email : Jacques.Moret-Bailly@u-bourgogne.fr
}}
\maketitle 

\medskip
Pacs 98.54.Aj Quasars 

\medskip

\abstract{The repartition of redshifts of the lines observed in the spectra of the quasars is generally 
considered as stochastic, but several authors showed that the difference of two redshifts is the product of 
an integer by a basic redshift $z_f = 0.062$.
This property results from a coherent Raman effect during the propagation of the incoherent light in a halo 
of atomic hydrogen, without any jet of gas, or dark matter. The coherence forbids a blur of the images or 
of the spectra.
The computation of $z_f$ does not require any new spectroscopic parameter. The non-linearity of the 
combination of Lyman absorptions and coherent Raman effect explains both the observed positions of the 
spectral lines and their high contrast.}

\section{Introduction}
The credibility of the standard explanations of the spectra of the quasars appears low: The Lyman lines of 
the atomic hydrogen appear with a lot of different redshifts generally considered as produced by a Doppler 
effect or by an expansion of the Universe. Both hypothesis have flaws because they require clouds of hot 
atomic hydrogen whose speed or stability cannot be explained using regular physics.

Studying the propagation of light in low pressure gases, one must take into account the \textquotedblleft 
Coherent Raman Effect on time-Incoherent Light\textquotedblright (CREIL) which transfers energy by 
frequency shifts from the hot modes of light to the thermal modes without any blur of the images and of the 
spectra; the relative frequency shifts $\Delta\nu/\nu$ due to CREIL are nearly constant.

In a previous paper \cite{Mor03}, we explained that the spectra can be obtained assuming that a nearly 
homogeneous cloud of Lyman pumped, atomic hydrogen is perturbed by a variable magnetic field: where 
the field is very low, there is nearly no redshift, the absorption (or emission) lines are written visibly into the 
spectrum; else, the redshift simultaneous with absorption (or emission) blurs the lines which become 
invisible. The spectrum results not from a modulation of the absorption, but from a modulation of the 
redshifting power of the gas. We showed the existence of a non-linearity able to increase the contrast of 
the lines.

However our previous explanation requires an important variation of the magnetic field for each line, thus, 
the existence of a large number of magnetised satellites. Worse, the spectra show a pseudo-stochastic 
repartition of the redshifts of the lines, the difference of two redshifts being the product of a constant $z_f = 
0.062$ by an integer \cite{Burbidge,Hewitt,Bell,Comeau}. It seems very difficult to 
explain such a spectroscopic regularity of the  \textquotedblleft intrinsic redshift\textquotedblright by the 
presence of objects on the line of sight; the non-linearity we had introduced provides a very simple 
solution.

\medskip
Section \ref{CREIL} reminds the key spectroscopic property which provides the intrinsic redshifts, the  
\textquotedblleft Coherent Raman Effect on time-Incoherent Light\textquotedblright  (CREIL) which is an 
avatar of the well known  \textquotedblleft Impulsive Stimulated Raman Scattering\textquotedblright  
(ISRS).

Section \ref{H} applies the CREIL to a halo of atomic hydrogen; to get the required resonances, H must 
have a non-zero orbital quantum number, so that the necessary Lyman pumping introduces a non-linearity.

\section{The  \textquotedblleft Coherent Raman Effect on time-Incoherent 
Light\textquotedblright as a limit case of the  \textquotedblleft Impulsive Stimulated Raman 
Scattering\textquotedblright.}\label{CREIL}

The  \textquotedblleft Impulsive Stimulated Raman Scattering\textquotedblright (ISRS) was discovered in 
1968 \cite{Giordmaine} and is commonly used to study the matter \cite{Yan,Weiner,Dougherty,Dhar}. 
ISRS is a light-matter interaction which transfers energy from a pulsed laser beam to a colder pulsed laser 
beam, by frequency shifts of the beams, the temperature of the beams being deduced from Planck's law. 
ISRS is space-coherent, that is the wave surfaces are left unchanged. ISRS is a parametric effect which 
uses matter as a catalyst, leaving it unchanged; it may be considered as a combination of two nearly 
simultaneous Raman scatterings giving virtual, opposite transitions. The relaxation times of the matter used 
as a catalyst must be longer that the length of the light pulses; these relaxation times are the period of (at 
least) a Raman-allowed transition, and, in a gas, the collisional time.

Replacing the ultrashort, powerful pulses of lasers by the pulses which make the ordinary incoherent light is 
a large change of the power and of the length of the pulses \cite{Mor98a,Mor98b,Mor01}.

The change of power has a qualitative effect: In ISRS, the Raman scattered amplitude which, by 
interference with the exciting amplitude, produces the frequency shifts, is an increasing function, nearly 
quadratic, of the exciting amplitude.
At low power, all scatterings are linear: If the coherence is broken by frequent collisions (ordinary Raman), 
the scattered intensity is proportional to the exciting intensity; else the coherently scattered amplitude is 
proportional to the exciting amplitude (Refraction, that is coherent Rayleigh scattering, Bragg scattering, 
coherent Raman scattering and photon echoes in microwaves\dots).
This behaviour which makes the frequency shifts independant on the intensities justifies a different name:  
\textquotedblleft Coherent Raman Effect on time-Incoherent Light\textquotedblright (CREIL). ISRS and 
CREIL may be computed from the tensors of polarisability of the molecules which do not depend much on 
the exciting frequencies, so that, in a first approximation, the relative frequency shift $\Delta\nu/\nu$ does 
not depend on the exciting frequency $\nu$.

As the length of the pulses, of the order of 5 nanoseconds, is larger than using lasers, the collisional time 
must be longer than 5 ns, so that the pressure must be lower than some Pascals. The Raman resonance 
period must be longer than 5 ns too, so that, to be active in CREIL, the molecules must have Raman 
resonances in the MHz range. Both conditions decrease the intensity of the frequency shift : therefore its 
observation requires a very long path which is not available in an usual laboratory.

In astrophysics, the thermal radiation, at least the 2.7K radiation, provides low temperature exciting beams 
whose blueshift corresponds to an amplification or a heating; the required long path is evidently available; it 
remains finding molecules having convenient Raman resonances. As hydrogen is the most common gas, 
two molecules may be considered:

- H$_2^+$ has Raman resonances close to 30 MHz, but it has probably not been observed although it 
should be produced by an ionisation of H$_2$ by the ultraviolet radiation. The reason is very simple: as 
H$_2^+$ is destroyed by the collisions with H$_2$ , its life-time is long only where the pressure is low 
enough to allow CREIL. A simultaneous absorption and CREIL makes absorption lines as wide as the 
redshift; therefore the almost forbidden lines of H$_2^+$ are so wide and weak that they cannot be seen.

- atomic hydrogen may have convenient transitions if it has been pumped by a Lyman transition so that 
hyperfine structures producing two photons resonances in the MHz range appear.

\section{Propagation of light in atomic hydrogen.} \label{H}

Consider the propagation of light having a continuous spectrum (constant intensity $I$, in particular in the 
Lyman region), in an homogeneous atmosphere of low pressure atomic hydrogen. In the fundamental state 
(principal quantum number n=1), the distance between the hyperfine levels (1420 MHz) is too large. In the 
other states, hyperfine transitions have convenient frequencies for the Raman allowed selection rule $\Delta 
F = 1$, for instance : 178 MHz in $2s_{1/2}$, 59 MHz in $2p_{1/2}$ and 24 MHz in $2p_{3/2}$.

Set $\Delta L$ the length of path for which the redshift is equal to the linewidth $\delta\nu$ of the Lyman 
$\alpha$ line, and assume that the atoms which are active in CREIL are mostly pumped by the Lyman 
$\alpha$ transition.

Set $\Delta\nu$ the redshift along the path $\Delta L$, which {\it would} result from a {\it complete} 
Lyman $\alpha$ absorption of the intensity $I$ , and suppose that, in a first approximation, the whole 
redshift results from the Lyman $\alpha$ absorption.

- case a: If $\Delta\nu$ is larger than the Lyman $\alpha$ linewidth $\delta\nu$ , that is if $I$ is large 
enough, $I$ is not fully absorbed, only the {\it constant} intensity $\Delta I$ which produces the redshift 
$\delta\nu$ is subtracted from $I$, that is from the spectrum while, by the redshift, the Lyman line crosses 
it. Thus, the contrast of lines which have been written into the spectrum is increased.

- case b: If, on the contrary, $\Delta\nu$ is lower than $\delta\nu$, the first approximation fails, a part of 
the redshift must result from other Lyman absorptions or other active atoms. Assuming a low redshifting 
power for these effects, a long path $\Delta L$ is necessary to get the redshift $\delta\nu$, so that the 
absorption of all lines is strong.

\medskip
If the intensity $I$ is constant and high, except for a single absorption line, the redshift and the absorption 
are constant (case a), except at a coincidence of the line with a Lyman line; at this coincidence, the 
redshifting power decreases (strongly if case b is reached), so that the absorption of {\bf all lines} of the 
gas is increased; similarly, a written emission line increases the redshifting power, so that the decrease of 
absorption appears as an emission; {\it the coincidence by redshift of a line already written in the spectrum 
with a Lyman line writes the whole spectral pattern of the gas into the spectrum.}

\section{The  \textquotedblleft forest\textquotedblright is made of  \textquotedblleft 
trees\textquotedblright.}\label{tree}
Suppose that a single Lyman pattern is written in the spectrum. The coincidence of the written, redshifted 
Lyman $\beta$ (resp. Lyman $\gamma$) line with the Lyman $\alpha$ line of the gas writes the Lyman 
pattern into the gas. Both patterns differ by the shift of frequencies $\nu_{(\beta {\rm resp.}\gamma)}-
\nu_\alpha$ of the $\alpha$ and $\beta$ (resp. $\gamma$) lines. As in the standard computations the lines 
are considered as Lyman $\alpha$, the frequency shift is relative to the Lyman $\alpha$ frequency:
\begin{equation}
z_{(\beta {\rm resp.}\gamma) , \alpha}=\frac{ \nu_{(\beta {\rm resp.}\gamma)}-\nu_\alpha}{ 
\nu_\alpha}\approx\frac{1-1/(3^2 {\rm resp. }4^2)-(1-1/2^2)}{1-1/2^2}
\end{equation}
\begin{equation}
z_{(\beta , \alpha)}\approx 5/27 \approx 0.1852 \approx 3*0.0617; z_{(\gamma , \alpha)}= 1/4 = 0.25 
=4*0.0625.
\end{equation}
Similarly $ z_{(\gamma , \beta)}\approx 7/108 \approx 0.065.$
The redshifts appear, with a good approximation as the products of $z_f = 0.062$ by an integer $q$.

The intensities of the Lyman lines are decreasing functions of the final principal quantum number $n$, so 
that the inscription of a pattern is better for $q =3$ than for $q = 4$ and {\it a fortiori} for $q = 1$.

\medskip
Iterating, the coincidences of the shifted lines frequencies with the Lyman $\beta$ or $\alpha$ frequencies 
build a tree, final values of $q$ being sums of the basic values 4, 3 and 1. Each step being characterised by 
the value of q, a generation of successive lines is characterised by successive values of $q: q_1, q_2...$ As 
the final redshift is $q_F*z_f = (q_1 +q_2 +...)*z_f $, the addition $q_F = q_1 + q_2 +...$ is both a 
symbolic representation of the successive elementary processes, and the result of these processes.

 The name  \textquotedblleft tree\textquotedblright is not very good because branches of the tree may be 
sticked by coincidences of frequencies. A remarkable coincidence happens for $q = 10$, this number 
being obtained by the effective coincidences deduced from:

\begin{equation}
10=3+3+4=3+4+3=4+3+3=3+3+3+1=...
\end{equation}
$q = 10$ is so remarkable that $z = 0.62$ may seem experimentally a value of $z$ more fundamental than 
$z_f$.
\medskip
In these computations, the levels for a value of the principal quantum number $n$ larger than 4 are not 
taken into account, assuming that the corresponding transitions are too weak.

Is the forest made of a single or several trees?

\section{Conclusion}
The present computation is a quantitative explanation of the frequencies observed in the spectra of the 
quasars. It requires only elementary physics, no unknown matter. The computation should be improved by 
an evaluation of the intensities.

The CREIL appears as a key in the study of the quasars. It should probably be helpful for other studies: 
for instance, some astrophysicists think that the dark matter needed to get the gravitational stability could 
be simply molecular hydrogen; if this is true, it exists some H$_2^+$, and a contribution of the CREIL not 
only to the  \textquotedblleft intrinsic\textquotedblright redshifts, but to the  \textquotedblleft 
cosmological\textquotedblright redshift.


\begin{thebibliography}{}%
\bibitem{Mor03} J. Moret-Bailly, 2003, paper submited to IEEE Transactions on Plasma Science; astro-
ph/0305180.
\bibitem{Burbidge}G. Burbidge, 1968 , {\it ApJ.}, {\bf 154}, L41
\bibitem{Hewitt}G. Burbidge \& A. Hewitt, 1990 , {\it ApJ.}, {\bf 359}, L33
\bibitem{Bell}M. B. Bell, 2002 , arXiv:astro-ph/0208320
\bibitem{Comeau}M. B. Bell \& S. P. Comeau, 2003 , arXiv:astro-ph/0305060
\bibitem{Giordmaine}Giordmaine, J. A. , M. A. Duguay \& J. W.Hansen, 1968, IEEE J. Quantum 
Electron., 4, 252
\bibitem{Yan}Yan, Y.-X. , E. B. Gamble Jr. \& K. A. Nelson , 1985,  J. Chem Phys., 83, 5391
\bibitem{Weiner}Weiner, A. M. , D. E. Leaird., G. P. Wiederrecht, \& K. A. Nelson,  1990, 
Science  247, 1317
\bibitem{Dougherty}Dougherty, T. P., G. P. Wiederrecht, K. A. Nelson, M. H. Garrett, H. P. 
Jenssen \& C. Warde, 1992, Science {\bf 258,}, 770
\bibitem{Dhar}Dhar, L. , J. A. Rogers, \& K. A. Nelson, 1994 Chem. Rev. {\bf 94}, 157 

\bibitem{Mor98a}J. Moret-Bailly, 1998, {\it Ann. Phys. Fr.}, {\bf 23}, C1-235.
\bibitem{Mor98b}J. Moret-Bailly, 1998, {\it Quantum and Semiclassical Optics}, {\bf 10}, L35.
\bibitem{Mor01}J. Moret-Bailly, 2001, {\it J. Quant. Spectr. \& rad. Transfer}, {\bf 68}, 575.

\end{thebibliography}
\end{document}